\documentclass[12pt]{article}
\usepackage{a4wide,epsfig,amsmath}

\allowdisplaybreaks

\newcommand{\agt}{\rlap{\lower 3.5 pt \hbox{$\mathchar \sim$}} \raise 1pt
 \hbox {$>$}}
\newcommand{\alt}{\rlap{\lower 3.5 pt \hbox{$\mathchar \sim$}} \raise 1pt
 \hbox {$<$}}
\newcommand{\Li}{\mathop{\mathrm{Li}}\nolimits}

\catcode`@=11
\newcount\@tempcntc
\def\@citex[#1]#2{\if@filesw\immediate\write\@auxout{\string\citation{#2}}\fi
  \@tempcnta\z@\@tempcntb\m@ne\def\@citea{}\@cite{\@for\@citeb:=#2\do
    {\@ifundefined
       {b@\@citeb}{\@citeo\@tempcntb\m@ne\@citea\def\@citea{,}{\bf
?}\@warning
       {Citation `\@citeb' on page \thepage \space undefined}}%
    {\setbox\z@\hbox{\global\@tempcntc0\csname b@\@citeb\endcsname\relax}%
     \ifnum\@tempcntc=\z@ \@citeo\@tempcntb\m@ne
       \@citea\def\@citea{,}\hbox{\csname b@\@citeb\endcsname}%
     \else
      \advance\@tempcntb\@ne
      \ifnum\@tempcntb=\@tempcntc
      \else\advance\@tempcntb\m@ne\@citeo
      \@tempcnta\@tempcntc\@tempcntb\@tempcntc\fi\fi}}\@citeo}{#1}}
\def\@citeo{\ifnum\@tempcnta>\@tempcntb\else\@citea\def\@citea{,}%
  \ifnum\@tempcnta=\@tempcntb\the\@tempcnta\else
   {\advance\@tempcnta\@ne\ifnum\@tempcnta=\@tempcntb \else
\def\@citea{--}\fi
    \advance\@tempcnta\m@ne\the\@tempcnta\@citea\the\@tempcntb}\fi\fi}
\catcode`@=12

\begin{document}

\title{
\vskip-3cm{\baselineskip14pt
\centerline{\normalsize DESY 20--214\hfill ISSN 0418-9833}
\centerline{\normalsize Dezember 2020\hfill}}
\vskip1.5cm
Angular analysis of bottom-flavored hadron production in semileptonic decays of
polarized top quarks}

\author{Bernd A. Kniehl\footnote{E-mail: kniehl@desy.de}\\
{\normalsize II. Institut f\"ur Theoretische Physik, Universit\"at Hamburg,}\\
{\normalsize Luruper Chaussee 149, 22761 Hamburg, Germany}\\
\\
S. Mohammad Moosavi Nejad\footnote{E-mail: mmoosavi@yazd.ac.ir}\\
{\normalsize Faculty of Physics, Yazd University, P.O. Box 89195--741, Yazd,
Iran}
}

\date{}

\maketitle

\begin{abstract}
We study the inclusive production of bottom-flavored hadrons from semileptonic
decays of polarized top quarks at next-to-leading order in QCD using
fragmentation functions recently determined from a global fit to $e^+e^-$ data.
We provide the relevant differential decay widths at parton level in analytic
form.
These results fill an important gap in the theoretical interpretation of
recent measurements of the top-quark polarization and the $t\bar{t}$ spin
correlations using dilepton final states in proton-proton collisions at the
CERN Large Hadron Collider.
As an illustration, we study the distributions in the scaled bottom-hadron
energy of the polarized-top-quark decay widths for different $W$-boson
helicities.
\medskip

\end{abstract}

\newpage

\section{Introduction}
\label{sec:one}

The top quark $t$ of the standard model (SM) is the heaviest known elementary
particle.
Due to its high mass, it plays a crucial role in testing the electroweak
symmetry breaking mechanism and in searching for new physics beyond the SM.
The precise determination of its properties, including its mass $m_t$ and total
decay width $\Gamma_t$, and its process-dependent features, like its
polarization or the correlation of its spin with that of a co-produced antitop
quark, is of prime importance.
The latter quantities are particularly sensitive probes of deviations from the
SM and allow us to constrain, e.g., the anomalous chromoelectric and
chromomagnetic dipole moments of the top quark.
The top-quark lifetime $\tau_t=\hbar/\Gamma_t\approx 5\times 10^{-25}$~s
\cite{Zyla:2020zbs} is shorter than the typical time scale of quantum
chromodynamics (QCD) $\hbar/\Lambda_\mathrm{QCD}\approx 10^{-24}$~s and much
shorter than the spin correlation time scale
$\hbar m_t/\Lambda_\mathrm{QCD}^2\approx 10^{-21}$~s, where
$\Lambda_\mathrm{QCD}$ is the asymptotic scale parameter of QCD.
Therefore, the top quark decays before it can hadronize, and its full spin
information is preserved during its decay process and fully encoded in the
angular distribution of its decay products.

The top-quark polarization and the $t\bar{t}$ spin correlations have recently
been measured using dilepton final states in Run~2 at the CERN Large Hadron
Collider (LHC) with center-of-mass energy $\sqrt{s}=13$~TeV, by the ATLAS
\cite{Aaboud:2019hwz} and CMS \cite{CMS:2018jcg} Collaborations.
Intriguingly, ATLAS found a deviation of 3.8 standard deviations ($\sigma$)
from the SM prediction of the asymmetry $A_{|\Delta\phi_{\ell\ell}|}$ of the
distribution in the azimuthal angular difference $\Delta\phi_{\ell\ell}$ of the
decay leptons.
This deviation was confirmed by CMS, albeit with a smaller significance of
about $2\sigma$.
These analyses are relying on the factorization of the squared matrix element
of the full process, 
$|\mathcal{M}(q\bar{q}/gg\to t\bar{t}\to b\ell^+\nu\bar{b}\ell^-\bar{\nu})|^2
\propto\rho\times R\times\bar{\rho}$, into the spin density matrices $R$,
$\rho$, and $\bar{\rho}$ for on-shell $t\bar{t}$ hadroproduction and
semileptonic $t$ and $\bar{t}$ decays, respectively, via the narrow-width
approximation.
While $R$ is treated in Refs.~\cite{Aaboud:2019hwz,CMS:2018jcg} at
next-to-leading order (NLO) in QCD
\cite{Bernreuther:2013aga,Bernreuther:2015yna},
$\rho$ and $\bar{\rho}$ are only modeled at leading order (LO) using the
program package {\sc madspin} \cite{Artoisenet:2012st}.
Moreover, the formation of bottom-flavored hadrons is not taken into account
within the rigorous framework of the QCD parton model with fragmentation
functions (FFs) whose universality is guaranteed by the factorization theorem
\cite{Collins:1998rz}.
It is an urgent matter to clarify in how far the observed deviations may be
related to a lack of precision in the theoretical treatment of the semileptonic
$t$ and $\bar{t}$ decays.

It is the purpose of the present paper to provide theoretical input needed to
fill this gap.
Specifically, we calculate the partial width of the inclusive decay\break
$t(\uparrow)\to b W^+(\uparrow)\to Bl^+\nu_l+X$, where $B$ generically denotes
a bottom-flavored hadron, at NLO in QCD allowing for top-quark polarization and
definite $W$-boson helicity and properly accounting for parton-to-hadron
fragmentation and finite-hadron-mass effects.
By doing so, we generalize our previous work \cite{Kniehl:2012mn}, where the
top-quark spin was averaged over.

The general-mass variable-flavor-number scheme (GM-VFNS), which has been
elaborated for inclusive heavy-flavored-hadron production in $e^+e^-$
annihilation \cite{Kneesch:2007ey}, two-photon collisions \cite{Kramer:2001gd},
photoproduction \cite{Kramer:2003jw}, and hadroproduction
\cite{Kniehl:2004fy,Kniehl:2005mk,Kniehl:2015fla,Benzke:2017yjn}, provides an
ideal theoretical framework also here.
However, owing to the large mass hierarchy $m_b\ll m_t$, finite-$m_b$
corrections are expected to be negligible in the case at hand.
This expectation was actually confirmed in Ref.~\cite{Kniehl:2012mn}, by a
comparative analysis of the partial width of the decay $t\to B+W^+$ in the
GM-VFNS and the zero-mass variable-flavor-number scheme (ZM-VFNS), where bottom
is included among the massless quark flavors.
In fact, the finite-$m_b$ corrections were found to be much smaller than the
contribution from gluon fragmentation.
Therefore, we will adopt the ZM-VFNS in the following.
However, we will include finite-$m_B$ effects, which modify the relations
between partonic and hadronic variables and reduce the available phase space, 
as explained in Sec.~2 of Ref.~\cite{Kniehl:2012mn}.

In Ref.~\cite{Kniehl:2012mn}, we adopted the $B$ FFs from
Ref.~\cite{Kniehl:2008zza}, which were determined at NLO in the ZM-VFNS
through a joint fit to $e^+e^-$ annihilation data taken by ALEPH
\cite{Heister:2001jg} and OPAL \cite{Abbiendi:2002vt} at CERN LEP1 and by SLD
\cite{Abe:1999ki} at SLAC SLC.
Specifically, the power ansatz $D_b(z,\mu_F^\text{ini})=Nz^\alpha(1-z)^\beta$
was used as the initial condition for the $b\to B$ FF at factorization scale
$\mu_F^\text{ini}=m_b=4.5$~GeV, while the gluon and light-quark FFs were
generated via the Dokshitzer-Gribov-Lipatov-Altatelli-Parisi (DGLAP)
\cite{Gribov:1972ri,Altarelli:1977zs,Dokshitzer:1977sg} evolution.
In Ref.~\cite{Salajegheh:2019ach}, the analysis of Ref.~\cite{Kniehl:2008zza}
was updated by including the data taken by DELPHI \cite{DELPHI:2011aa} at CERN
LEP1, which were published after Ref.~\cite{Kniehl:2008zza}, working both at
NLO and next-to-next-to-leading order (NNLO) with the same theoretical assumptions.
In our numerical analysis, we employ the new $B$ FFs from
Ref.~\cite{Salajegheh:2019ach}.

A similar analysis, albeit without fragmentation and finite-$m_B$ effects, was
reported in Refs.~\cite{Fischer:1998gsa,Fischer:2001gp}.
Our analysis provides an independent check of analytic results presented
therein.
A related NLO analysis with a different treatment of the final state, for
bottom jets instead of bottom hadrons, was performed in
Ref.~\cite{Bernreuther:2014dla}, leading to results that, unfortunately, cannot
be compared with ours in any straightforward way.
Recently, the analysis of Refs.~\cite{Fischer:1998gsa,Fischer:2001gp} has been
extended to NNLO in QCD using the optical
theorem \cite{Czarnecki:2018vwh}.
Due to the totally inclusive treatment of the hadronic part of the final state,
this result does not allow for the implementation of FFs.

This paper is organized as follows.
In Sec.~\ref{sec:two}, we list our parton-level results in analytic form,
relegating lengthy formulas to the Appendix.
In Sec.~\ref{sec:three}, we present our numerical analysis.
In Sec.~\ref{sec:four}, we summarize our conclusions.

\section{Analytic results}
\label{sec:two}

We work at NLO in the ZM-VFNS, implemented in the modified minimal-subtraction
($\overline{\mathrm{MS}}$) scheme, and consider the decay process
\begin{equation}
t(p_t,s)\to b(p_b)+W^+(p_W,\lambda)({}+g(p_g))
\to B(p_B)+\ell^+(p_\ell)+\nu_\ell(p_\nu)+X,
\label{eq:proc}
\end{equation}
where $X$ collectively denotes the unobserved final-state hadrons and the
four-momentum, spin, and helicity assignments are indicated in parentheses.
We have $s=\pm1/2$ and $\lambda=0,\pm1$.
The gluon in Eq.~\eqref{eq:proc} contributes to the real radiation at NLO.
Both the $b$ quark and the gluon may hadronize into the $B$ hadron.
For simplicity, we employ the narrow-width approximation, where $p_W^2=m_W^2$
and small terms of order $\mathcal{O}(\Gamma_W^2/m_W^2)$ are neglected.
As mentioned in Sec.~\ref{sec:one}, we put $m_b=0$, but keep $m_B$ finite.
In the top-quark rest frame, the $b$ quark, gluon, and $B$ hadron have energies
$E_i=p_t\cdot p_i/m_t$ ($i=b,g,B$), which nominally range from
$E_b^\text{min}=E_g^\text{min}=0$ and $E_B^\text{min}=m_B$ to
$E_b^\text{max}=E_g^\text{max}=(m_t^2-m_W^2)/(2m_t)$ and
$E_B^\text{max}=(m_t^2+m_B^2-m_W^2)/(2m_t)$, respectively.
As in Ref.~\cite{Kniehl:2012mn}, we choose the scaling variable $z$ by setting
$E_B=zE_a$, with $a=b,g$, in the range $0\le z\le1$.
Introducing the scaled energies $x_i=E_i/E_b^\text{max}$ ($i=b,g,B$), we then
have $m_B/E_b^\text{max}\le x_B\le x_a\le1$.
An alternative definition of the scaling variable, in terms of light-cone
variables, is discussed in Sec.~3 of Ref.~\cite{Kniehl:2012mn}, to where we
refer the interested reader.

\begin{figure}[t]
\begin{center}
\includegraphics[width=0.75\linewidth]{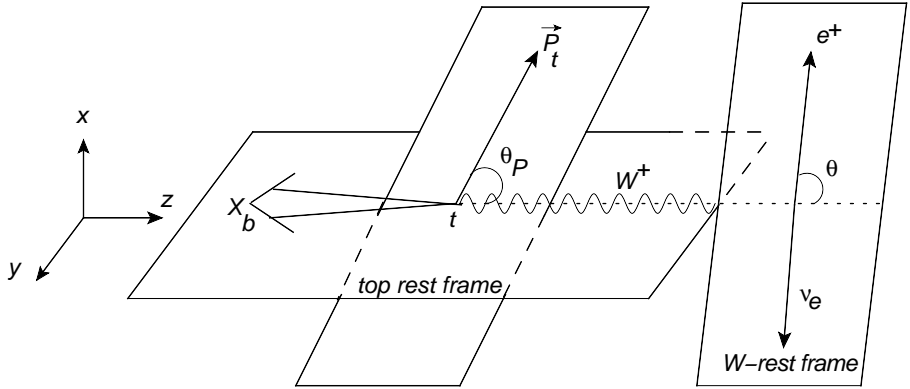}
\caption{\label{fig:angle}%
Definitions of the polar angle $\theta_P$ of the top-quark polarization
three-vector in the top-quark rest frame and of the polar angle $\theta$ of the
charged-lepton three-momentum in the $W$-boson rest frame.
In both cases, the $z$ axis is chosen to point along the $W$-boson
three-momentum in the top-quark rest frame.}
\end{center}
\end{figure}

We implement the polarization of the top quark by writing its (average) spin
four-vector in its rest frame as
$s_t^\mu=P(0,\sin\theta_P\cos\phi_P,\sin\theta_P\sin\phi_P,\cos\theta_P)$,
where $P$ is the magnitude of the top-quark polarization, taking values in the
range $0\le P\le1$.
Here, it is understood that the $z$ axis is chosen to point along the $W$-boson
flight direction, so that $s_t\cdot p_W=-P|\vec{p}_W|\cos\theta_P$.
To describe the leptonic decay of the $W$ boson, we boost into the rest frame
of the latter, which leaves the $z$ axis invariant, and define the
charged-lepton four-momentum to be
$p_\ell^\mu=E_\ell(1,\sin\theta\cos\phi,\sin\theta\sin\phi,\cos\theta)$.
The polar angles $\theta_P$ and $\theta$, which appear in our final results,
are also illustrated in Fig.~\ref{fig:angle}.

We wish to calculate the triply differential partial decay width 
$d^3\Gamma/(dx_B\,d\cos\theta\,d\cos\theta_P)$ of process~\eqref{eq:proc}.
Analogously to Eq.~(3) in Ref.~\cite{Kniehl:2012mn}, we have
\begin{equation}
\frac{d^3\Gamma}{dx_B\,d\cos\theta\,d\cos\theta_P}
=\sum_{a=b,g}\int_{x_B}^1\frac{dx_a}{x_a}\,
\frac{d^3\hat{\Gamma}_a}{dx_a\,d\cos\theta\,d\cos\theta_P}(\mu_R,\mu_F)
D_a\left(\frac{x_B}{x_a},\mu_F\right),
\label{eq:master}
\end{equation}
where $\mu_R$ is the renormalization scale, $\mu_R$ is the factorization scale,
$D_a(z,\mu_F)$ is the $a\to B$ FF, and
\begin{equation}
\frac{d^3\hat{\Gamma}_a}{dx_a\,d\cos\theta\,d\cos\theta_P}
=\frac{1}{2}\left(\frac{d^2\hat{\Gamma}_a^\text{unpol}}{dx_a\,d\cos\theta}
+P\frac{d^2\hat{\Gamma}_a^\text{pol}}{dx_a\,d\cos\theta}\cos\theta_P\right).
\label{eq:thetap}
\end{equation}
The factor $1/2$ on the right-hand side of Eq.~\eqref{eq:thetap} ensures that
\begin{equation}
\int_{-1}^1d\cos\theta_P\,
\frac{d^3\hat{\Gamma}}{dx_a\,d\cos\theta\,d\cos\theta_P}
=\frac{d^2\hat{\Gamma}_a^\text{unpol}}{dx_a\,d\cos\theta}.
\end{equation}
Each $W$-boson helicity is featured by a characteristic $\theta$ dependence,
which is encoded in the structure
\begin{equation}
\frac{d^2\hat{\Gamma}_a}{dx_a\,d\cos\theta}
=\frac{3}{8}(1+\cos\theta)^2\frac{d\hat{\Gamma}_a^{+}}{dx_a}
+\frac{3}{8}(1-\cos\theta)^2\frac{d\hat{\Gamma}_a^{-}}{dx_a}
+\frac{3}{4}\sin^2\theta\frac{d\hat{\Gamma}_a^{0}}{dx_a}.
\label{eq:theta}
\end{equation}
This holds for both the unpolarized and polarized terms in
Eq.~\eqref{eq:thetap}.
Notice that the $\theta$-dependent coefficients on the right-hand side of
Eq.~\eqref{eq:theta} are normalized so that, upon integration over
$\cos\theta$, we have
\begin{equation}
\frac{d\hat{\Gamma}_a}{dx_a}=\sum_{\lambda=-1}^1
\frac{d\hat{\Gamma}_a^\lambda}{dx_a}.
\label{eq:int}
\end{equation}

At LO, we only have $a=b$, and $x_b=1$ is fixed, i.e.\ the $x_b$ dependence
comes as a delta-function peak.
Specifically, we have
\begin{eqnarray}
\hat{\Gamma}_{b,\text{LO}}^{0,\text{unpol}}&=&\hat\Gamma_{b,\text{LO}}^{0,\text{pol}}=
F(1-\omega)^2,
\nonumber\\
\hat{\Gamma}_{b,\text{LO}}^{-,\text{unpol}}&=&-\hat\Gamma_{b,\text{LO}}^{-,\text{pol}}=
F(2\omega)(1-\omega)^2,
\nonumber\\
\hat{\Gamma}_{b,\text{LO}}^{+,\text{unpol}}&=&\hat\Gamma_{b,\text{LO}}^{+,\text{pol}}=0,
\label{eq:born}
\end{eqnarray}
where $\omega=m_W^2/m_t^2$ and
$F=G_Fm_t^3|V_{tb}|^2B(W^+\to\ell^+\nu_\ell)/(8\pi\sqrt{2})$.
Here, $G_F$ is Fermi's constant, $V_{ij}$ is the $ij$ element of the
Cabibbo-Kobayashi-Matrix quark mixing matrix
\cite{Cabibbo:1963yz,Kobayashi:1973fv}, and $B(W^+\to\ell^+\nu_\ell)$ is
the branching ratio of the leptonic $W$-boson decay mode considered.
Neglecting the masses of the charged leptons and the first five quark flavors,
we have
\begin{eqnarray}
  B(W^+\to\ell^+\nu_\ell)&=&
  \frac{1}{3+N_c\sum_{\substack{i=u,c\\ j=d,s,b}}|V_{ij}|^2
    \left[1+3C_F\alpha_s(\mu_R)/(4\pi)\right]}
  \nonumber\\
  &\approx&\frac{1}{9}\left[1-\frac{2}{3}\,\frac{\alpha_s(\mu_R)}{\pi}\right],
\label{eq:br}
\end{eqnarray}  
where we have included the NLO QCD correction, with color factors $N_c=3$ and
$C_F=(N_c^2-1)/(2N_c)=4/3$. 
In the last equality in Eq.~\eqref{eq:br}, we have approximated
$V_{ij}\approx\delta_{ij}$.
At LO, the top-quark spin is passed on to the bottom quark for $\lambda=0$,
while it is flipped for $\lambda=-1$;
the case $\lambda=+1$ is forbidden by angular-momentum conservation in the
limit $m_b\to0$, as is reflected in Eq.~\eqref{eq:born}.
Using Eq.~\eqref{eq:int}, we have
\begin{eqnarray}
\hat{\Gamma}_{b,\text{LO}}^\text{unpol}
&=&F(1+2\omega)(1-\omega)^2,\nonumber\\
\hat{\Gamma}_{b,\text{LO}}^\text{pol}
&=&F(1-2\omega)(1-\omega)^2.
\end{eqnarray}
Inserting Eq.~\eqref{eq:born} in Eq.~\eqref{eq:master}, we obtain a very simple
formula for the final LO result:
\begin{eqnarray}
  \frac{d^3\Gamma_\text{LO}}{dx_B\,d\cos\theta\,d\cos\theta_P}
  &=&\frac{3}{8}F(1-\omega)^2D_b(x_B,\mu_F)
  [\sin\theta(1+P\cos\theta_P)
    \nonumber\\
&&{}+\omega(1-\cos\theta)^2(1-P\cos\theta_P)].
\end{eqnarray}  

\begin{figure}[t]
\begin{center}
\includegraphics[width=0.75\linewidth]{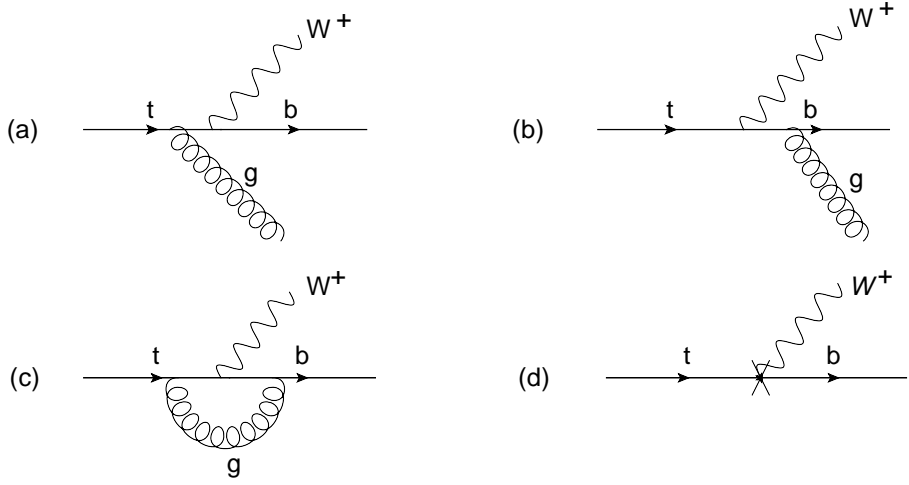}
\caption{\label{fig:feynman}%
Feynman diagrams contributing to the partial decay width of the process in
Eq.~\eqref{eq:proc} at NLO:
(a) initial-state radiation;
(b) final-state radiation;
(c) vertex correction; and
(d) combination of wave function renormalizations and vertex counterterm.
The leptonic $W$-boson decay is not shown.}
\end{center}
\end{figure}

The Feynman diagrams contributing to the partial decay width of the process in
Eq.~\eqref{eq:proc} at NLO are depicted in Fig.~\ref{fig:feynman}.
The NLO coefficient functions of the unpolarized case may be found in
Appendix~A of Ref.~\cite{Kniehl:2012mn} and those of the polarized case are
presented in the Appendix of this paper.
At NLO, the case $\lambda=+1$ is enabled by the presence of the additional
spin-one gluon even for $m_b=0$.

At this point, we compare our new analytic results with the literature.
In Refs.~\cite{Fischer:1998gsa,Fischer:2001gp}, process~\eqref{eq:proc} was
also considered at NLO using the narrow-width approximation and putting
$m_b=0$, but treating the final state in a less differential fashion, which
does not allow for the convolution with $a\to B$ FFs on the basis of the
ZM-VFNS.
We can compare our results for $\hat{\Gamma}_b^{\lambda,\text{pol}}$ with
Refs.~\cite{Fischer:1998gsa,Fischer:2001gp} upon integration over $x_b$ in
the range $0\le x_b\le1$.
Specifically, the quantities $\hat{\Gamma}_U^P$, $\hat{\Gamma}_L^P$, and
$\hat{\Gamma}_F^P$ listed in Eqs.~(18)--(20) of Ref.~\cite{Fischer:1998gsa}
(see also Eqs.~(42)--(44) in Ref.~\cite{Fischer:2001gp}) are related to
$\hat{\Gamma}_b^{\lambda,\text{pol}}$ as
$\hat{\Gamma}_b^{0,\text{pol}}/\hat{\Gamma}_{b,\text{LO}}^\text{unpol}
=\hat{\Gamma}_L^P$ and
$\hat{\Gamma}_b^{\pm,\text{pol}}/\hat{\Gamma}_{b,\text{LO}}^\text{unpol}
=(\hat{\Gamma}_U^P\pm\hat{\Gamma}_F^P)/2$.
With these identifications, we fully agree with
Refs.~\cite{Fischer:1998gsa,Fischer:2001gp}.

\section{Numerical results}
\label{sec:three}

We are now in a position to explore the phenomenological consequences of our
results by performing a numerical analysis.
We adopt from Ref.~\cite{Zyla:2020zbs} the input parameter values
$G_F = 1.1663787\times10^{-5}$~GeV$^{-2}$,
$m_W = 80.379$~GeV,
$m_t = 172.4$~GeV,
$m_B = 5.279$~GeV,
$|V_{tb}|=1$, and
$B(W^+\to\ell^+\nu_\ell)=10.86\%$.
We evaluate $\alpha_s^{(n_f)}(\mu_R)$ at NLO (NNLO) in the
$\overline{\text{MS}}$ scheme using Eq.~(4) of Ref.~\cite{Kniehl:2006bg},
retaining only the first two (three) terms on the right-hand side, with $n_f=5$
active quark flavors and asymptotic scale parameter
$\Lambda_{\overline{\text{MS}}}^{(5)}=225$~MeV (207~MeV) adjusted such that
$\alpha_s^{(5)}(m_Z) = 0.1179$ for $m_Z = 91.1876$~GeV \cite{Zyla:2020zbs}.
This yields $\alpha_s^{(5)}(m_t) = 0.1076$ (0.1076).
As already mentioned in Sec.~\ref{sec:one}, we use the up-to-date $B$ FFs from
Ref.~\cite{Salajegheh:2019ach}, both at NLO and NNLO.
For definiteness, we identify $\mu_R=\mu_F=\xi m_t$ and vary $\xi$ from
1/2 to 2 about the default value 1 to estimate the theoretical uncertainty due
to the lack of knowledge of higher-order corrections.

The angular dependencies at the parton level, in Eqs.~\eqref{eq:thetap} and
\eqref{eq:theta}, are passed on to the hadron level via
Eq.~\eqref{eq:master}.
The hadron level counterparts, $d\Gamma^\lambda/dx_B$, of the coefficient
functions $d\hat{\Gamma}_a^\lambda/dx_a$ in Eq.~\eqref{eq:theta} may be
projected out from the measured $\theta$ distribution
$d\Gamma/(dx_B\,d\cos\theta)$ as explained in Sec.~4 of
Ref.~\cite{Kniehl:2012mn} (see Eqs.~(21)--(23) therein) and thus represent
physical observables by themselves.
In Ref.~\cite{Kniehl:2012mn}, the top quarks were taken to be unpolarized,
i.e., according to our present notation,
$d\Gamma^{\lambda,\text{unpol}}/dx_B$ were considered.
In the following, we complement the study of Ref.~\cite{Kniehl:2012mn} by
presenting predictions for $d\Gamma^{\lambda,\text{pol}}/dx_B$.
For the sake of a coherent treatment, we also provide the analogous predictions
for $d\Gamma^{\lambda,\text{unpol}}/dx_B$, thus updating the analysis of
Ref.~\cite{Kniehl:2012mn}.

Our central predictions are of NLO.
To assess their significance, we compare them with the respective LO results.
Our LO predictions are slightly inconsistent because they are evaluated with
NLO FFs.
Unfortunately, Ref.~\cite{Salajegheh:2019ach} does not provide a LO FF set.
By the way, the same is true for Ref.~\cite{Kniehl:2008zza}, to which we could
have resorted otherwise.
On the other hand, Ref.~\cite{Salajegheh:2019ach} also supplies a NNLO set.
While consistent NNLO predictions are unfeasible in the absence of NNLO
parton-level results, this still offers us the opportunity to get a first
impression of the typical magnitude of the NNLO effects.
In our pseudo-NNLO analysis, besides using NNLO FFs, we also evaluate
$\alpha_s(\mu_R)$ at NNLO as explained above.

\begin{figure*}[t]
\begin{center}
\includegraphics[width=0.480\textwidth]{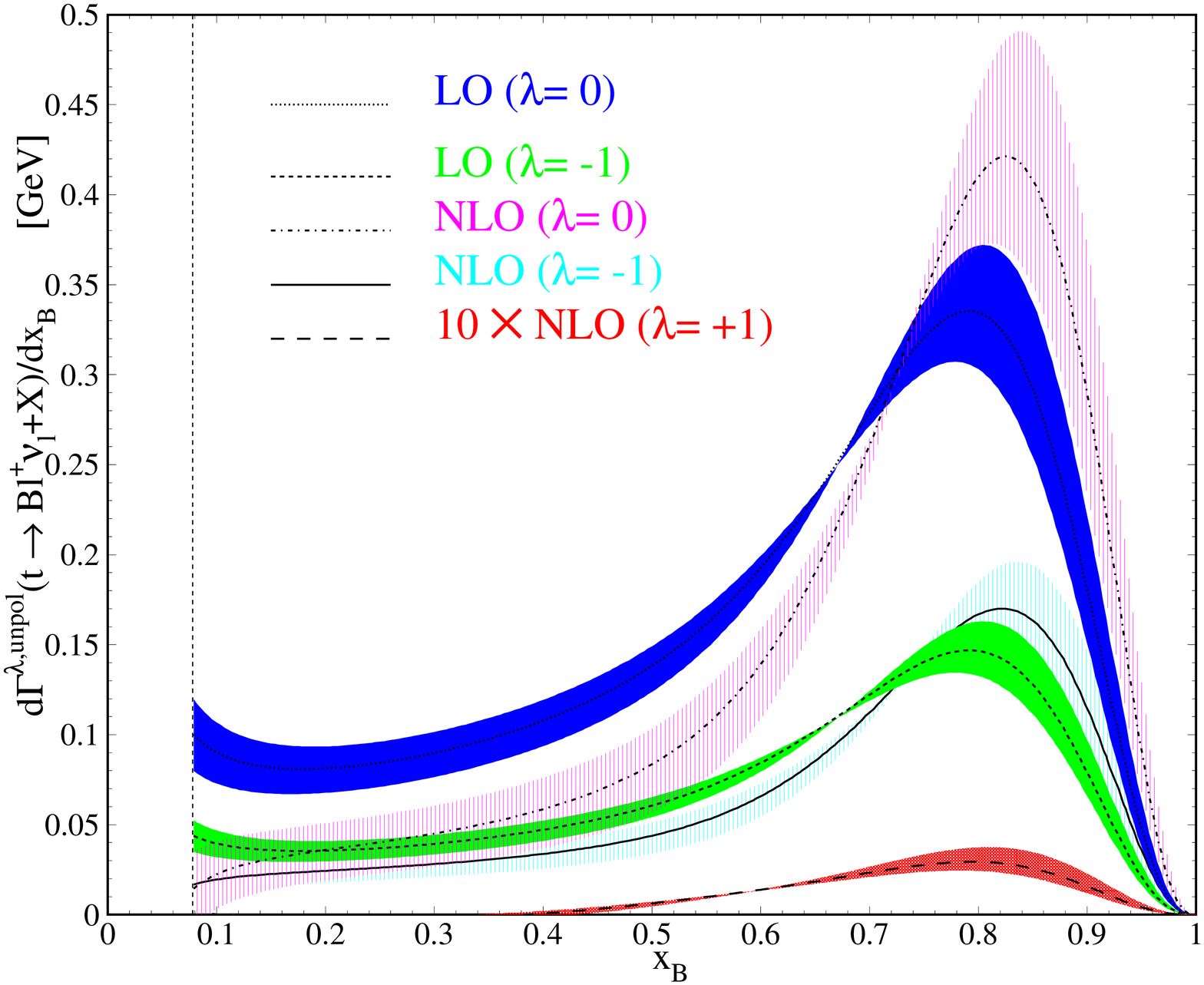}
\includegraphics[width=0.480\textwidth]{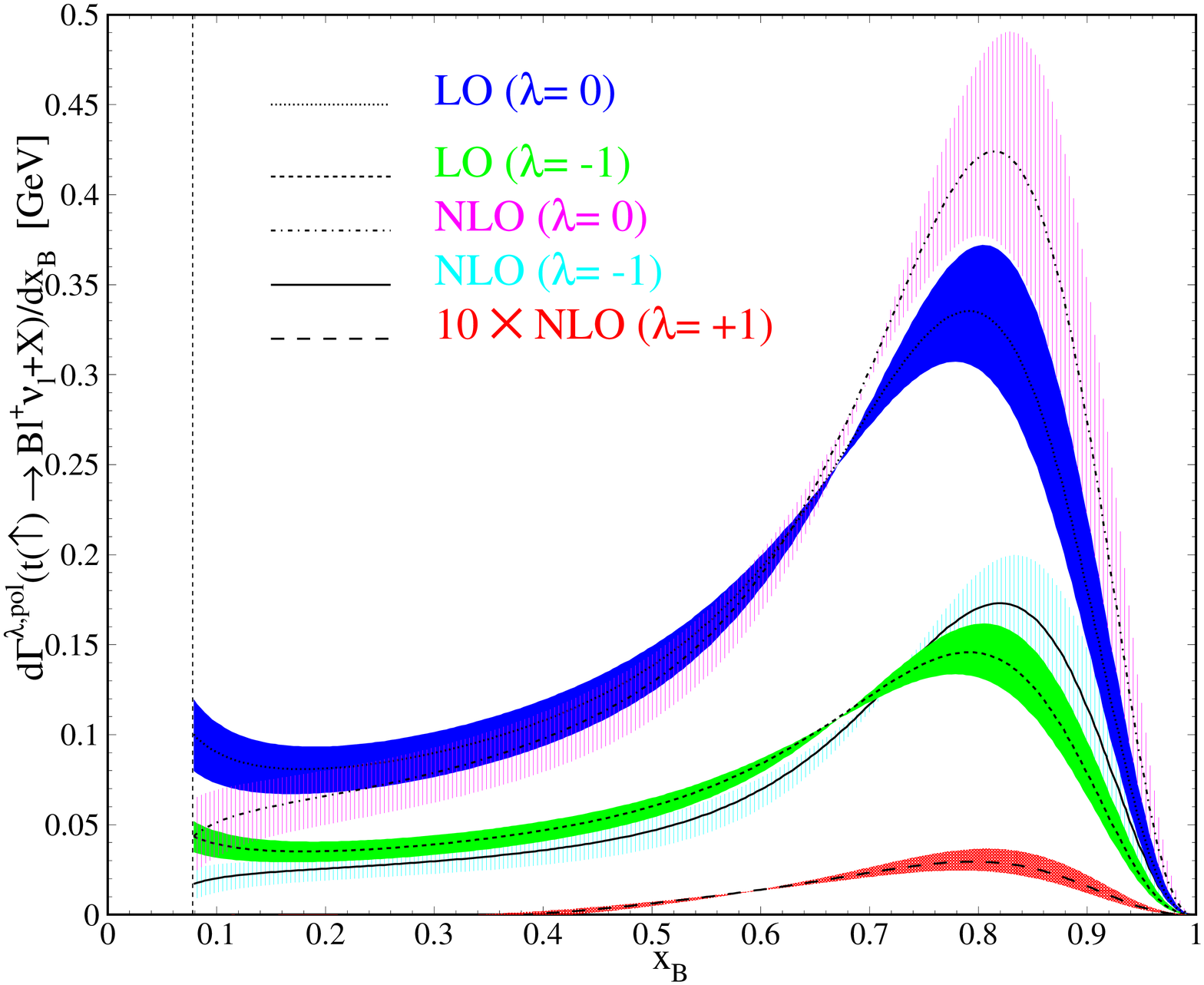}
\caption{\label{fig:two}
  LO and NLO results for (a) $d\Gamma^{\lambda,\text{unpol}}/dx_B$ and (b)
  $d\Gamma^{\lambda,\text{pol}}/dx_B$ with $\lambda=0,\pm1$ as functions of
  $x_B$.
  $d\Gamma_{\text{NLO}}^{+,\text{unpol}}/dx_B$ and
  $d\Gamma_{\text{NLO}}^{+,\text{pol}}/dx_B$ are rescaled by a factor of 10 for
  better visibility.
  The theoretical uncertainties are indicated by the shaded bands.}
\end{center}
\end{figure*}

\begin{figure*}[t]
\begin{center}
\includegraphics[width=0.480\textwidth]{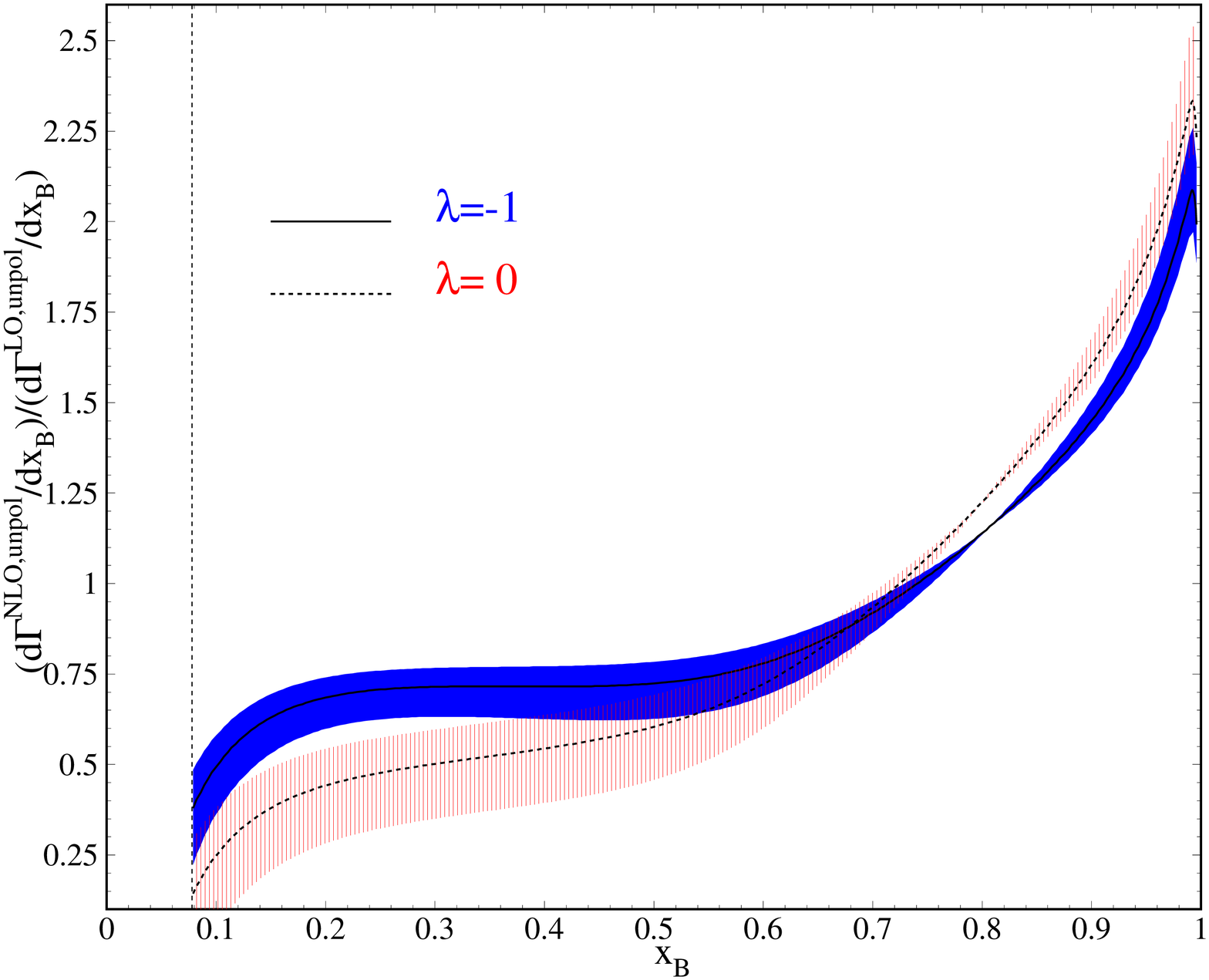}
\includegraphics[width=0.480\textwidth]{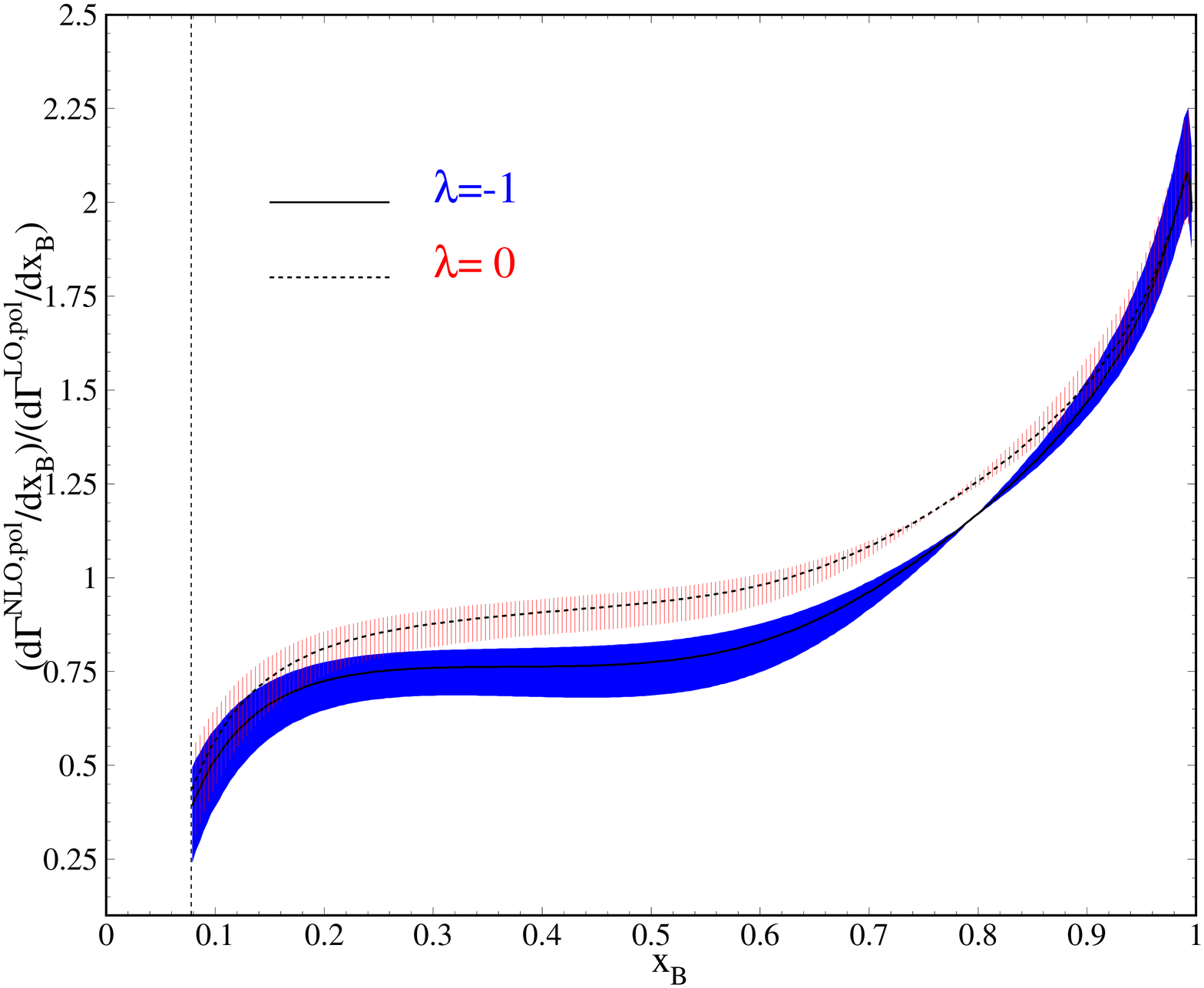}
\caption{\label{fig:three}
  NLO results for (a) $d\Gamma^{\lambda,\text{unpol}}/dx_B$ and (b)
  $d\Gamma^{\lambda,\text{pol}}/dx_B$ with $\lambda=0,-1$,
  normalized to the respective default LO results, as functions of $x_B$.
  The theoretical uncertainties of the NLO results are indicated by the shaded
  bands.}
\end{center}
\end{figure*}

In Fig.~\ref{fig:two}, we study the $x_B$ dependencies of
$d\Gamma^{\lambda,\text{unpol}}/dx_B$ and $d\Gamma^{\lambda,\text{pol}}/dx_B$
for $\lambda=0,\pm1$ at NLO and compare them with the respective LO results for
$\lambda=0,-1$.
As explained in Sec.~\ref{sec:two}, $d\Gamma^{+,\text{unpol}}/dx_B$ and
$d\Gamma^{+,\text{pol}}/dx_B$ vanish at LO in our approximation.
The theoretical uncertainties are indicated by the shaded bands.
The slight inconsistency in our LO analysis mentioned above, not only affects
the default predictions, but also the estimation of the theoretical
uncertainty, which is expected to be slightly larger if the $\mu_F$ dependence
is subject to LO DGLAP evolution.
Comparing Figs.~\ref{fig:two}(a) and (b), we observe that
$d\Gamma^{\lambda,\text{unpol}}/dx_B$ and $d\Gamma^{\lambda,\text{pol}}/dx_B$
are very similar as for normalization and line shape.
Each $x_B$ distribution exhibits a maximum close to $x_B=0.8$.
Longitudinal $W$-boson helicity is favored, with
$d\Gamma_{\text{LO}/\text{NLO}}^{0,\text{unpol}/\text{pol}}/dx_B$ being more
than twice as large as
$d\Gamma_{\text{LO}/\text{NLO}}^{-,\text{unpol}/\text{pol}}/dx_B$
for negative $W$-boson helicity $\lambda=-1$.
On the other hand, positive $W$-boson helicity $\lambda=+1$ is perturbatively
suppressed; rescaling $d\Gamma_{\text{NLO}}^{+,\text{unpol}/\text{pol}}/dx_B$
by the inverse couplant, $2\pi/\alpha_s^{(5)}(m_t)$, brings it up to the level
of $d\Gamma_{\text{NLO}}^{-,\text{unpol}/\text{pol}}/dx_B$.
The NLO corrections have a significant effect on the $x_B$ distributions for
$\lambda=0,-1$, by raising their peaks and lowering their small-$x_B$ tails.
To render these features more visible, we present, in Fig.~\ref{fig:three}, the
QCD correction ($K$) factors for $\lambda=0,-1$, which we evaluate by
normalizing the NLO predictions including their theoretical-uncertainty bands
relative to the default LO predictions.
From Fig.~\ref{fig:three}(a) and (b), we observe that, both for unpolarized and
polarized top quarks, the $K$ factors steadily increase by one order of
magnitude, typically from 0.2 to 2, as $x_B$ runs across its range of values.
We conclude from Figs.~\ref{fig:two} and \ref{fig:three} that the NLO
corrections are quite significant and should be taken into account in
theoretical interpretations of future top-quark polarization measurements.

\begin{figure*}[t]
\begin{center}
\includegraphics[width=0.480\textwidth]{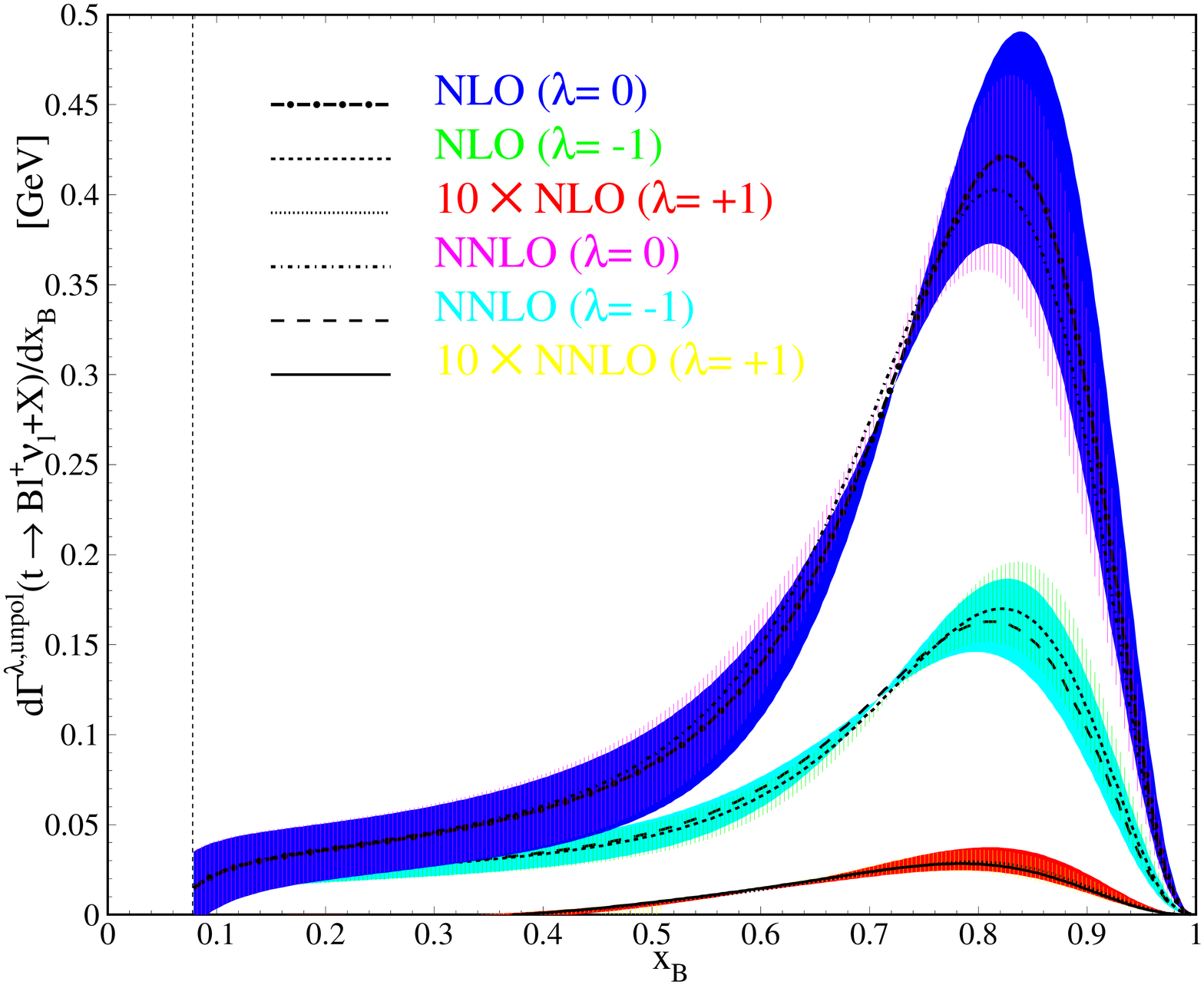}
\includegraphics[width=0.480\textwidth]{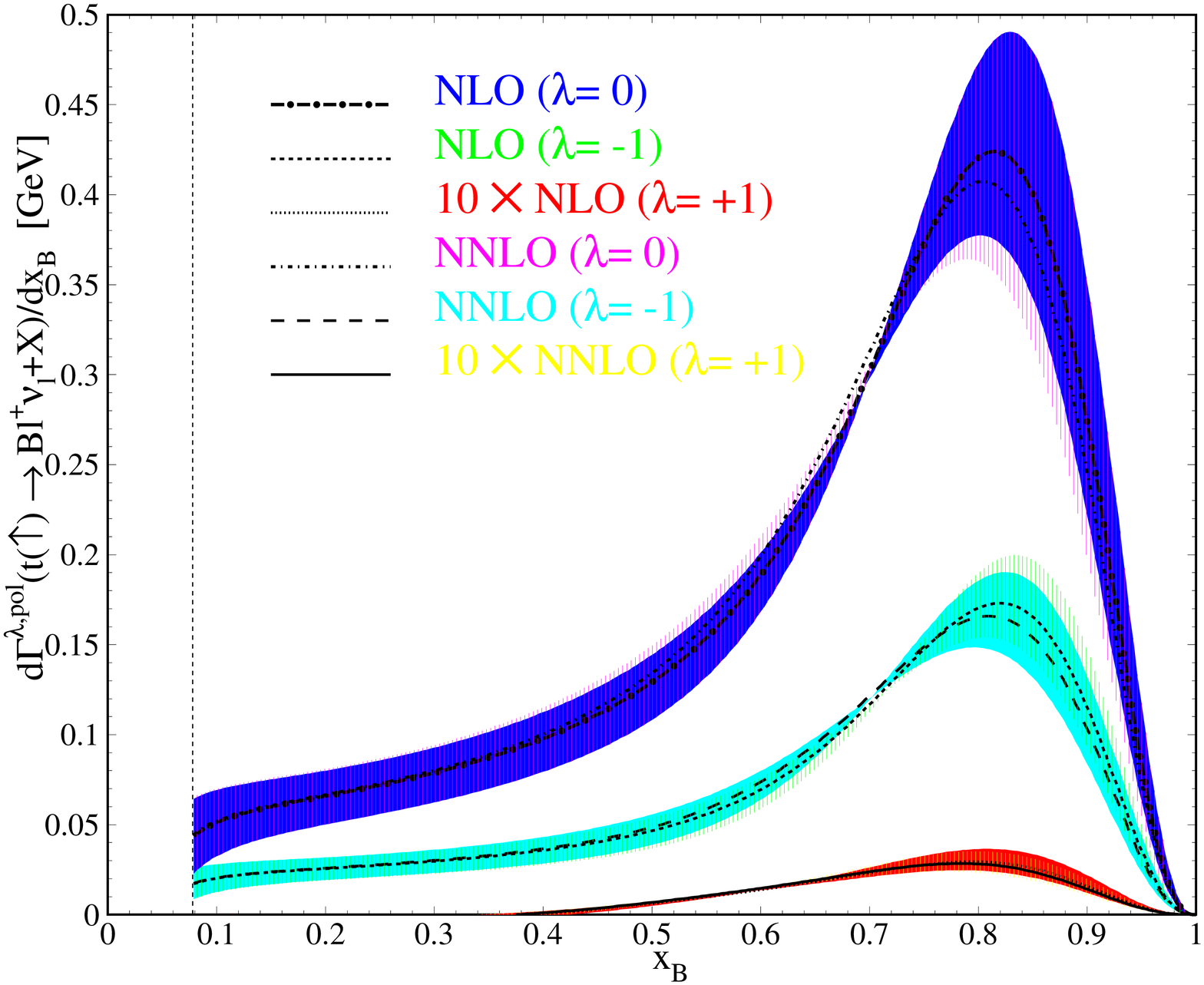}
\caption{\label{fig:four}
  NLO and pseudo-NNLO results for (a) $d\Gamma^{\lambda,\text{unpol}}/dx_B$ and
  (b) $d\Gamma^{\lambda,\text{pol}}/dx_B$  with $\lambda=0,\pm1$ as functions
  of $x_B$.
  $d\Gamma_{\text{NLO}/\text{NNLO}}^{+,\text{unpol}}/dx_B$ and
  $d\Gamma_{\text{NLO}/\text{NNLO}}^{+,\text{pol}}/dx_B$ are rescaled by a
  factor of 10 for better visibility.
  The theoretical uncertainties are indicated by the shaded bands.}
\end{center}
\end{figure*}

\begin{figure*}[t]
\begin{center}
\includegraphics[width=0.480\textwidth]{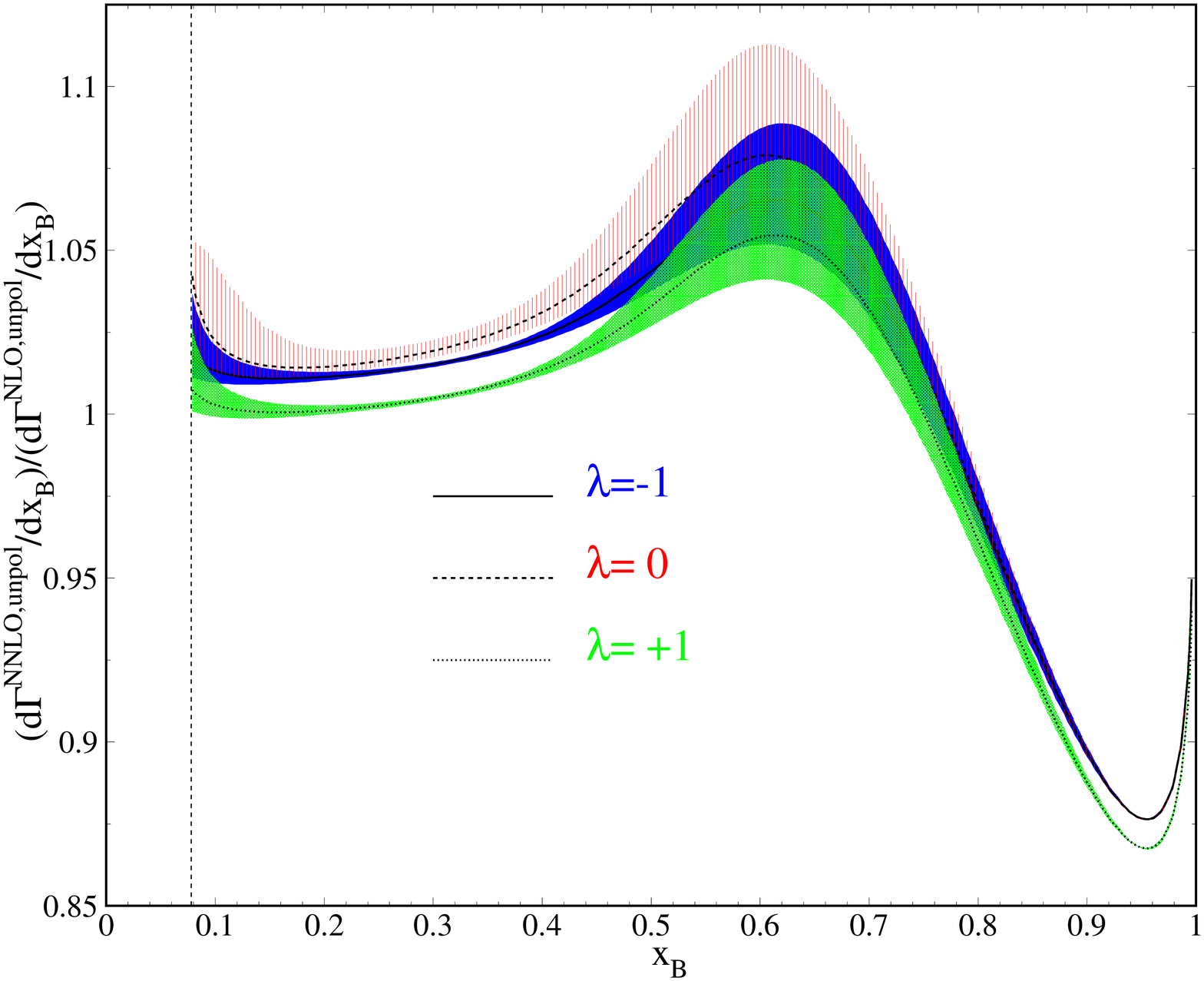}
\includegraphics[width=0.480\textwidth]{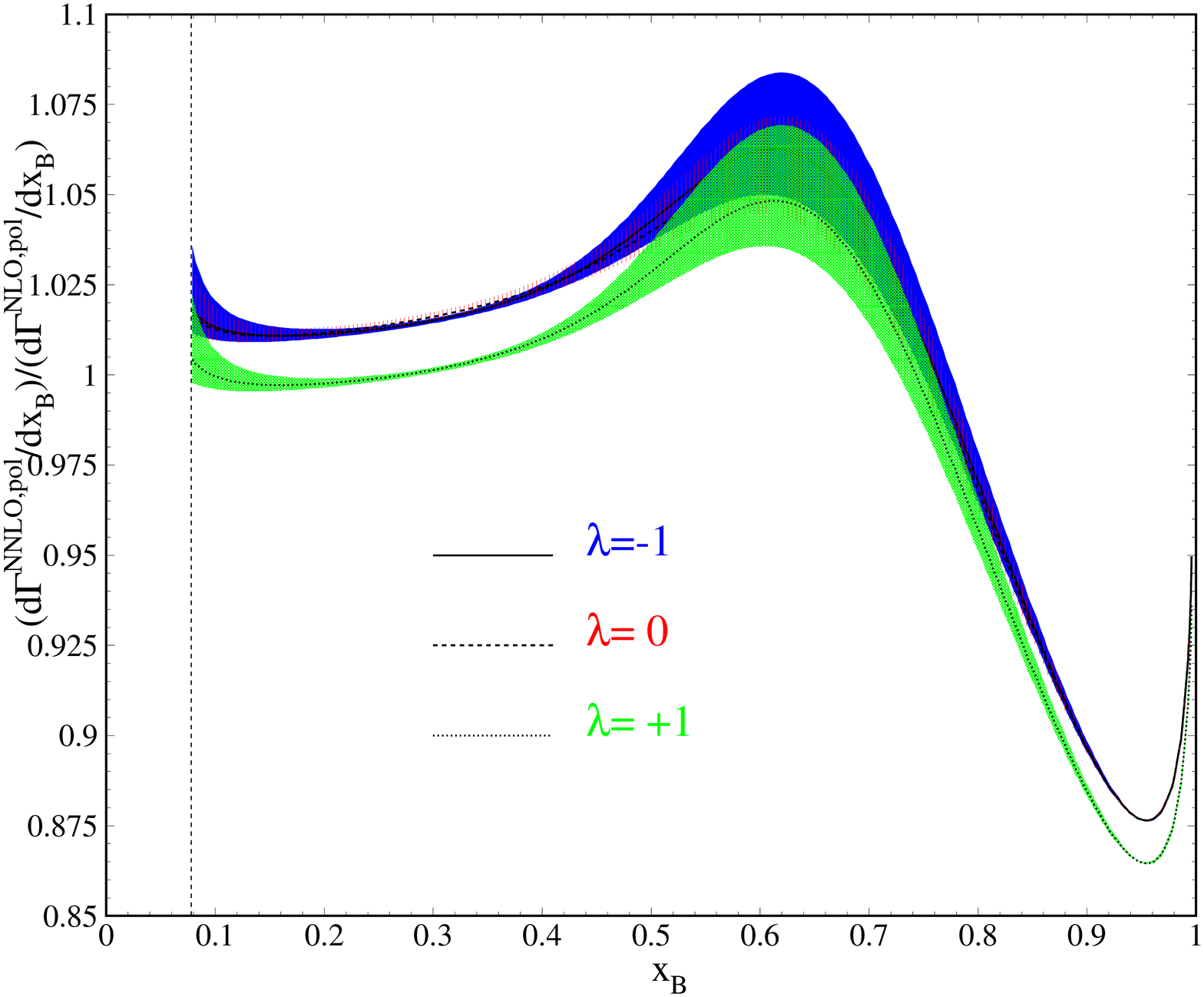}
\caption{\label{fig:five}
  Pseudo-NNLO results for (a) $d\Gamma^{\lambda,\text{unpol}}/dx_B$ and (b)
  $d\Gamma^{\lambda,\text{pol}}/dx_B$ with $\lambda=0,\pm1$, normalized to the
  respective default NLO results, as functions of $x_B$.
  The theoretical uncertainties of the pseudo-NNLO results are indicated by the
  shaded bands.}
\end{center}
\end{figure*}

In Fig.~\ref{fig:four}, compare our pseudo-NNLO evaluations of
$d\Gamma^{\lambda,\text{unpol}}/dx_B$ and $d\Gamma^{\lambda,\text{pol}}/dx_B$
for $\lambda=0,\pm1$ with the respective NLO results already presented in
Fig.~\ref{fig:two}.
For better visibility, we also plot, in Fig.~\ref{fig:five}, the ratios of
the pseudo-NNLO results including their theoretical-uncertainty bands and the
default NLO predictions.
We conclude from Figs.~\ref{fig:four} and \ref{fig:five} that the NNLO effects
are likely to be relatively modest, of the order of 10\% or less.

\section{Summary and Conclusions} 
\label{sec:four}

We studied the inclusive production of bottom-flavored hadrons from
semileptonic decays of polarized top quarks at NLO in the ZM-FVNS using the
narrow-width approximation for the intermediate $W$ bosons, whose
helicities we distinguished.
Specifically, we considered the partial decay width differential in the
scaled bottom-hadron energy $x_B$, the azimuthal angle $\theta$ of the charged
lepton in the $W$-boson rest frame, and the azimuthal angle $\theta_P$ of the
top-quark polarization in the top-quark rest frame.
In our numerical analysis, we employed up-to-date $B$ FFs, recently determined
from a global fit to all available $e^+e^-$ data \cite{Salajegheh:2019ach}.
We thus extended our previous study in Ref.~\cite{Kniehl:2012mn}, which was
restricted to unpolarized top quarks.
In Ref.~\cite{Kniehl:2012mn}, we had convinced ourselves, by comparing
evaluations in the GM-VFNS and ZM-VFNS, that finite-$m_b$ corrections may be
safely neglected.
On the other hand, we retained finite-$m_B$ corrections, which reduce the
available phase space and lead to visible effects at small values of $x_B$.
We provided full analytic results, ready to be used by the interested reader.

We found the NLO QCD corrections to be quite significant, inducing a reduction
in the lower $x_B$ range and an enhancement in the upper $x_B$ range, with $K$
factors ranging from 0.2 to 2.
On the other hand, including partial information from NNLO, contained in the
evaluation of the $B$ FFs and $\alpha_s(\mu_R)$, turned out to yield only mild
modifications, with the due caveat that such results suffer from a violation of
renormalization group invariance already in the considered order.

Our combined results, from Ref.~\cite{Kniehl:2012mn} and this paper, help us to
fill an important gap in the theoretical interpretation of recent measurements
of the top-quark polarization and the $t\bar{t}$ spin correlations in dilepton
final states at the LHC \cite{Aaboud:2019hwz,CMS:2018jcg}.
In particular, it will be interesting to see if these theoretical improvements
will contribute to reconciling the ATLAS~\cite{Aaboud:2019hwz} and
CMS~\cite{CMS:2018jcg} measurements of $A_{|\Delta\phi_{\ell\ell}|}$ with the
SM expectations.

\section*{Acknowledgments}

We thank G.~Kramer for useful discussions at the initial stage of this work
and C.~Schwanenberger for detailed information on Ref.~\cite{CMS:2018jcg}.
The work of B.A.K. was supported in part by the German Federal Ministry for
Education and Research BMBF through Grant No.\ 05H18GUCC1.
The work of S.M.M.N. was supported in part by the Iran National Science
Foundation INSF through Grant No.\ 97005414.

\section*{Appendix}

In this appendix, we list the coefficient functions
$d\hat{\Gamma}_a^{\lambda,\text{pol}}/dx_a$ appearing in
Eq.~\eqref{eq:theta} at NLO in the ZM-VFNS with $m_b=0$.
They possess the following structure:
\begin{eqnarray}
\frac{1}{\hat{\Gamma}_{b,\text{LO}}^\text{pol}}\,
\frac{d\hat{\Gamma}_a^{0,\text{pol}}}{dx_a}
&=&\frac{1}{1-2\omega}\left[\delta_{ab}\delta(1-x_a)
+\frac{\alpha_s(\mu_R)}{2\pi}\left(P_{ab}(x_a)\ln\frac{m_t^2}{\mu_F^2}
+C_FC_a^{0,\text{pol}}(x_a)\right)\right],
\nonumber\\
\frac{1}{\hat{\Gamma}_{b,\text{LO}}^\text{pol}}\,
\frac{d\hat{\Gamma}_a^{-,\text{pol}}}{dx_a}
&=&\frac{-2\omega}{1-2\omega}\left[\delta_{ab}\delta(1-x_a)
+\frac{\alpha_s(\mu_R)}{2\pi}\left(P_{ab}(x_a)\ln\frac{m_t^2}{\mu_F^2}
+C_FC_a^{-,\text{pol}}(x_a)\right)\right],
\nonumber\\
\frac{1}{\hat{\Gamma}_{b,\text{LO}}^\text{pol}}\,
\frac{d\hat{\Gamma}_a^{+,\text{pol}}}{dx_a}
&=&\frac{4\omega}{(1-2\omega)(1-\omega)^2}
\frac{\alpha_s(\mu_R)}{2\pi}C_FC_a^{+,\text{pol}}(x_a),
\label{eq:structure}
\end{eqnarray}
where
\begin{eqnarray}
P_{qq}(x)&=&C_F\left(\frac{1+x^2}{1-x}\right)_+,\nonumber\\
P_{gq}(x)&=&C_F\frac{1+(1-x)^2}{x}
\end{eqnarray}
are the timelike $q\to q$ and $q\to g$ splitting functions at LO.

For $a=b$, we have

{\scriptsize
\begin{eqnarray}
C_b^{0,\text{pol}}(x)&=&
-\delta(1-x)\big[2\ln\omega\ln(1-\omega)+4\Li_2(\omega)+\frac{2\omega}{1-\omega}\ln\omega+6\big]+2(1+x^2)\bigg(\frac{\ln(1-x)}{1-x}\bigg)_+\nonumber\\
&&{}+2\frac{1+x^2}{(1-x)_+}\ln(x(1-\omega))+\frac{8(x-2)(1-x)^2}{x^2(Sx^2-2 x+2)}+2\omega(1-x)-\frac{12}{x}+\frac{8}{x^2}+\frac{6}{1-x}-\frac{(1+x)^2}{1-x}-\frac{1+x^2}{1-x}R_1\nonumber\\
&&{}+\bigg(1-|1-2x+2Sx^2|\bigg)\bigg(6\frac{1-x}{x^2}-\frac{1}{(1-x)(1-\omega)}+6\frac{(1-x)^2(x-2)^2}{x^2(Sx^2-2x+2)^2}+\frac{1}{2x^2(Sx^2-2x+2)}\big[9x^3\nonumber\\
&&{}-57x^2+97x-\frac{1}{1-x}-47\big]\bigg)+\frac{R_2}{S^{3/2}(Sx^2-2x+2)^{5/2}}\bigg(x^3S^4(-x^3+7x^2-22x+10+\frac{2}{1-x})+xS^3(x^4\nonumber\\
&&{}+4x^3+25x^2-32x+12)-2(1-x)^2+S^2(1-x)(5x^3+21x^2-12x+4)-S(1-x)(9x^2-4x-4)\bigg),
\label{eq:bz}\\
C_b^{-,\text{pol}}(x)&=&-\delta(1-x)\big[2\ln\omega\ln(1-\omega)+4\Li_2(\omega)+\frac{2\omega}{1-\omega}\ln\omega +6+\frac{1-\omega}{\omega}\ln(1-\omega)\big]+2(1+x^2)\bigg(\frac{\ln(1-x)}{1-x}\bigg)_+\nonumber\\
&&{}+2\frac{1+x^2}{(1-x)_+}\ln(x(1-\omega))-\frac{1+x^2}{2(1-x)}R_1+\frac{(1-S)^2}{\omega S}+\frac{\omega+S x}{\omega}-\frac{\omega}{2S(1-x)}+\frac{(1-S-Sx)}{S}\ln(1-2Sx)
\nonumber\\
&&{}-\frac{3S(2x-1)+6(1-x)}{2S(Sx^2-2x+2)^2}+\frac{S^2(1+3x)+4x\omega-3S}{2\omega S(Sx^2-2x+2)}+B_3|1-2x+2Sx^2|
+B_1\ln|2Sx^2-2x+1|\nonumber\\
&&{}+\frac{R_3}{S^2\sqrt{\omega}}B_2-\frac{R_2}{2\sqrt{Sx^2-2x+2}}\bigg(x^2\sqrt{S}+\frac{(S-1)x}{\sqrt{S}}-\frac{2\sqrt{S}}{1-x}+\frac{2S^2-5S+7}{S^{\frac{3}{2}}}\nonumber\\
&&{}+\frac{Sx(4S^2-23S+16)-S(28-19S)+6}{S^{\frac{5}{2}}(Sx^2-2x+2)}-\frac{3Sx(12-9S+S^2)+15S(S-2)+12(1-x)}{S^{\frac{5}{2}}(Sx^2-2x+2)^2}\bigg),
\label{eq:bm}\\
C_b^{+,\text{pol}}(x)&=&\frac{(1+x^2)R_1}{1-x}S^2-\frac{2(1+x)S^3}{\omega}+2S\big[1+Sx+(1-S-Sx)\ln(1-2Sx)\big]\nonumber\\
&&{}-\frac{(17-21 x)S^3+(20x-27)S^2+4(2-x)S}{\omega(Sx^2-2x+2)}+\frac{6(1-x)S-3(1-2x)S^2}{(Sx^2-2x+2)^2}+\frac{\omega S}{1-x}+\frac{2R_3}{\sqrt{\omega}}B_2\nonumber\\
&&{}-2S^2B_3|1-2x+2Sx^2|+2S^2 B_1 \ln|2Sx^2-2x+1|-\frac{R_2(1-x)\sqrt{S}}{x^3\sqrt{Sx^2-2x+2}}\bigg(11+7x^2-16x\nonumber\\
&&{}+\frac{6(2-x)(1-x)^3}{(2-2x+Sx^2)^2}+\frac{16x^3-58x^2+69x-28}{2-2x+Sx^2}+\frac{Sx^2(x^2+5x-4)}{1-x}+\frac{S^2x^4(1+x^2)}{(1-x)^2}\bigg),
\label{eq:bp}
\end{eqnarray}
}
where $S=(1-\omega)/2$,
\begin{eqnarray}
R_1&=&\ln\bigg[(1-S)x^2-x+\frac{1}{2}+\frac{1}{2}|2Sx^2-2x+1|\bigg],\nonumber\\
R_2&=&\ln\bigg[(1-x)(1-3Sx)+Sx^2(1-2Sx)+
\sqrt{S(Sx^2-2x+2)}|2Sx^2-2x+1|\bigg]-\nonumber\\
&&\ln\bigg[1+(S-1)x+\sqrt{S(Sx^2-2x+2)}\bigg],\nonumber\\
R_3&=&\ln\big[1-x(1-\sqrt{\omega})\big]-\ln\big|1-x(1+\sqrt{\omega})|,\nonumber\\
B_1&=&x-\frac{1-S}{S}-\frac{1-x}{2S(2-2x+Sx^2)^2}\big[5S^2x^2-2(1-x)-2S(x^2+3x-3)\big],\nonumber\\
B_2&=&-\omega S+xS^2
+\frac{Sx(-4S^2+7S-2)+S(8-10S)-2}{2(Sx^2-2x+2)}-\frac{\omega(S^2 x-4Sx+3S+2x-2)}{(Sx^2-2x+2)^2},\nonumber\\
B_3&=&\frac{1}{2S(1-x)}+\frac{1}{1-2Sx}+\frac{6(1-x)-3S(1-2x)}{2S(Sx^2-2x+2)^2}+\frac{S^2(3x-5)+S(11-2x)-4}{2\omega S(Sx^2-2x+2)}.
\end{eqnarray}

For $a=g$, we have
{\scriptsize
\begin{eqnarray}
C_g^{0,\text{pol}}(x)&=&\frac{1+(1-x)^2}{x}\bigg(-R_1+2\ln[x(1-\omega)(1-x)]\bigg)+\bigg(8S-x-26+\frac{13}{S}+\frac{26S^2-38S+13}{2S^3x^2}\nonumber\\
&&{}+\frac{-44S^2+88S-35}{2xS^2}\bigg)R_4+\bigg(\frac{1-6S}{2(2Sx-1)}+9\frac{1-S}{2Sx}+\frac{-4S^2+25S-13}{2S^2x^2}+\frac{\omega}{2(1-2Sx)^2}\bigg)|2Sx^2-2x+1|\nonumber\\
&&{}+2(7-6S)+(1-4S)x+7\frac{7S-5}{2Sx}+\frac{4S^2-25S+13}{2S^2x^2},
\label{eq:gz}\\
C_g^{-,\text{pol}}(x)&=&\frac{1+(1-x)^2}{2x}\bigg(4\ln[x(1-\omega)(1-x)]-R_1\bigg)+\frac{1-6S}{4\omega}x-\frac{\omega^2}{32S^3x^2(1-2Sx)^2}-\frac{(10S-7)\omega}{32S^3x^2(1-2Sx)}\nonumber\\
&&{}+\frac{1+5S-8S^2}{2S\omega}-A_3R_4+\frac{|2Sx^2-2x+1|}{8\omega S^2 x^2}A_5+\frac{A_2}{2S^3x^2}\ln(1-2Sx)+\frac{R_3 A_4}{2\sqrt{\omega}S^3x^2}-A_1\frac{\ln|2Sx^2-2x+1|}{2x^2S^3}\nonumber\\
&&{}+\frac{1}{16\omega x S^2}(48S^3+72S^2-50S-5)+\frac{1}{16\omega S^3x^2}(32S^3-88S^2+44S-3),
\label{eq:gm}\\
C_g^{+,\text{pol}}(x)&=&\frac{1}{2S x^2}\Big[\frac{2R_3}{\sqrt{\omega}}A_4+2x(1+(1-x)^2)S^3R_1+4 S^3 x^2 A_3R_4+2A_2\ln(1-2Sx)-2A_1\ln|2Sx^2-2x+1|\nonumber\\
&&{}-\frac{SA_5}{2\omega}|2Sx^2-2x+1|+\frac{S}{4\omega}\bigg(4S^2(1+2S)x^3+8S(4S^2-7S+1)x^2+\frac{8\omega^2 x}{S(1-4S)}-(96S^2-70S+5)x\nonumber\\
&&{}+\frac{32S^2-12S-3}{S}-\frac{40S^3-68S^2+38S-7}{2S(1-2Sx)}+\frac{8S^3-12S^2+6S-1}{2S(1-2Sx)^2}\bigg)\Big],
\label{eq:gp}
\end{eqnarray}
}
where
\begin{eqnarray}
R_4&=&\ln\bigg[1-S(-2Sx^2+2x+1-|2Sx^2-2x+1|)\bigg]-\ln[1-2Sx],\nonumber\\
A_1&=&2x^2(x-1)S^3+(-2x^2+5x+1)S^2-2(2+x)S+2,\nonumber\\
A_2&=&x(x^2-2)S^3+(-2x^2+5x+1)S^2-2(2+x)S+2,\nonumber\\
A_3&=&\frac{2+x^2}{2x}-\frac{5x^2+5x+1}{2Sx^2}-\frac{7-S(8+15x)}{4x^2S^3},\nonumber\\
A_4&=&2x(2-x^2)S^3+(2x^2-7x-4)S^2+2(x+3)S-2,\nonumber\\
A_5&=&20S-11-Sx(10S-7)-\frac{\omega(7-10S)}{2(1-2Sx)}+\frac{\omega^2}{2(1-2Sx)^2}.
\end{eqnarray}

Adding Eqs.~\eqref{eq:bz}--\eqref{eq:bp} for $a=b$ and
Eqs.~\eqref{eq:gz}--\eqref{eq:gp} for $a=g$ according to Eq.~\eqref{eq:int}, we
obtain the respective coefficient functions pertaining to the case where
$\theta$ is integrated over.
Specifically, we have
\begin{eqnarray}
\frac{1}{\hat{\Gamma}_{b,\text{LO}}^\text{pol}}\,
\frac{d\hat{\Gamma}_b^\text{pol}}{dx_b}
&=&\delta(1-x_b)+\frac{\alpha_s(\mu_R)}{2\pi}
\left(P_{qq}(x_b)\ln\frac{m_t^2}{\mu_F^2}+C_FC_b^\text{pol}(x_b)\right),
\nonumber\\
\frac{1}{\hat{\Gamma}_{b,\text{LO}}^\text{pol}}\,
\frac{d\hat{\Gamma}_g^\text{pol}}{dx_g}
&=&\frac{\alpha_s(\mu_R)}{2\pi}
\left(P_{gq}(x_g)\ln\frac{m_t^2}{\mu_F^2}+C_FC_g^\text{pol}(x_g)\right),
\end{eqnarray}
where
{\scriptsize
\begin{eqnarray}
C_b^\text{pol}(x)&=&\delta(1-x)\big[-2\ln\omega\ln(1-\omega)+\frac{2(1-\omega)}{1-2\omega}\ln(1-\omega)-4\Li_2(\omega)-\frac{2\omega}{1-\omega}\ln\omega-6\big]\nonumber\\
&&{}+2(1+x^2)\bigg(\frac{\ln(1-x)}{1-x}\bigg)_++2\frac{1+x^2}{(1-x)_+}\ln(x(1-\omega))+\frac{1-S}{S(Sx^2-2 x+2)}-\frac{2}{1-4S}-\frac{\omega}{S(1-x)}-1-x\nonumber\\
&&{}-\frac{1+x^2}{1-x}R_1+\bigg(\frac{1}{S(1-x)}+\frac{4\omega}{(1-4S)(1-2Sx)}-\frac{1-S}{S(Sx^2-2x+2)}\bigg)|2Sx^2-2x+1|\nonumber\\
&&{}-\frac{R_2}{\sqrt{S(Sx^2-2x+2)}}\bigg(\frac{1}{S}+\frac{2}{1-4S}+\frac{(13S-5-4S^2)x}{1-4S}-\frac{Sx(x^2-x+2)}{1-x}-\frac{(1-S)[1-(1-S)x]}{S(Sx^2-2x+2)}\bigg),\nonumber\\
C_g^\text{pol}(x)&=&
\frac{1+(1-x)^2}{x}\bigg(-R_1+2\ln[x(1-\omega)(1-x)]\bigg)-\bigg(x-\frac{1+\omega^2}{4S^3x^2}-\frac{8S^2-6S+3}{S(4S-1)}-\frac{2-\omega^2(2\omega-5)}{2S^2(1-4S)x}\bigg)R_4\nonumber\\
&&{}+\bigg(\frac{S-1}{2S^2x^2}+\frac{\omega}{2(1-2Sx)^2}-\frac{12S^2-15S+7}{2S(1-4S)x}-\frac{24S^2-26S+9}{2(1-4S)(1-2Sx)}\bigg)|2Sx^2-2x+1|+x+\frac{1-S}{2S^2x^2}\nonumber\\
&&{}-\frac{2}{1-4S}+\frac{12S^2-7S+5}{2S(1-4S)x}.
\end{eqnarray}
}

\end{document}